\documentclass[sigconf]{acmart}
\usepackage{natbib}

\usepackage{xspace}
\usepackage{multirow}
\usepackage{url}

\usepackage{algorithm}
\usepackage{algorithmic}

\newcommand{\kernelname}{DA-SpMM\xspace}

\newcommand{\modified}[1]{\textcolor{black}{#1}}

\usepackage[normalem]{ulem}
\newif\ifmodify
\modifytrue % Uncomment to compile full version

\ifmodify
\newcommand{\dl}[1]{\textcolor{red}{\sout{#1}}}
\newcommand{\guyue}[1]{\textcolor{orange}{#1}}
\newcommand{\comments}[1]{( {#1} )}
\else
\newcommand{\dl}[1]{}
\newcommand{\guyue}[1]{#1}
\newcommand{\comments}[1]{\iffalse {#1} \fi}
\fi

\newcommand{\squishlist}{
   \begin{list}{$\bullet$}
    { \setlength{\itemsep}{0pt}      \setlength{\parsep}{0pt}
      \setlength{\topsep}{0pt}       \setlength{\partopsep}{0pt}
      \setlength{\listparindent}{-2pt}
      \setlength{\itemindent}{-5pt} 
      \setlength{\leftmargin}{1em} \setlength{\labelwidth}{0em}
      \setlength{\labelsep}{0.5em} } }
\newcommand{\squishend}{
    \end{list}  }

\renewcommand\footnotetextcopyrightpermission[1]{}
\AtBeginDocument{%
  \providecommand\BibTeX{{%
    \normalfont B\kern-0.5em{\scshape i\kern-0.25em b}\kern-0.8em\TeX}}}

\setcopyright{none}
\begin{document}

%%
%% The "title" command has an optional parameter,
%% allowing the author to define a "short title" to be used in page headers.
\title{Heuristic Adaptability to Input Dynamics for SpMM on GPUs}

\author{Guohao Dai$^1$*, Guyue Huang$^2$, Shang Yang$^1$, Zhongming Yu$^1$, Hengrui Zhang$^1$}
\author{Yufei Ding$^2$, Yuan Xie$^2$, Huazhong Yang$^1$, Yu Wang$^1$*}
\author{$^1$Tsinghua University, $^2$University of California at Santa Barbara}
\author{*Corresponding authors: daiguohao@mail.tsinghua.edu.cn, yu-wang@tsinghua.edu.cn}

\renewcommand{\shortauthors}{Guohao Dai \textit{et al.}}

\begin{abstract}

    Sparse Matrix-Matrix Multiplication (SpMM) has served as fundamental components in various domains. Many previous studies exploit GPUs for SpMM acceleration because GPUs provide high bandwidth and parallelism. We point out that a static design does not always improve the performance of SpMM on different input data (\textit{e.g.,} >85\% performance loss with a single algorithm). In this paper, we consider the challenge of input dynamics from a novel auto\guyue{-}tuning perspective, while following issues remain to be solved:
    \textbf{(1) Orthogonal design principles considering sparsity.} Orthogonal design principles for such a sparse problem should be extracted to form different algorithms, and further used for performance tuning. \textbf{(2) Nontrivial implementations in the algorithm space.} Combining orthogonal design principles to create new algorithms needs to tackle with new challenges like thread race handling. \textbf{(3) Heuristic adaptability to input dynamics.} The heuristic adaptability is required to dynamically optimize code for input dynamics.

    To tackle these challenges, we first propose a novel three-loop model to extract orthogonal design principles for SpMM on GPUs. The model not only covers previous SpMM designs, but also comes up with new designs absent from previous studies. We propose techniques like conditional reduction to implement algorithms missing in previous studies. We further propose DA-SpMM, a Data-Aware heuristic GPU kernel for SpMM. DA-SpMM adaptively optimizes code considering input dynamics. Extensive experimental results show that, DA-SpMM achieves \textbf{1.26$\times$}$\sim$\textbf{1.37$\times$} speedup compared with the best NVIDIA cuSPARSE algorithm on average, and brings up to \textbf{5.59$\times$} end-to-end speedup to applications like Graph Neural Networks.

\end{abstract}

\maketitle

\section{Introduction}

\begin{figure}[!bp]
    \centering
    \includegraphics[width=0.48\textwidth]{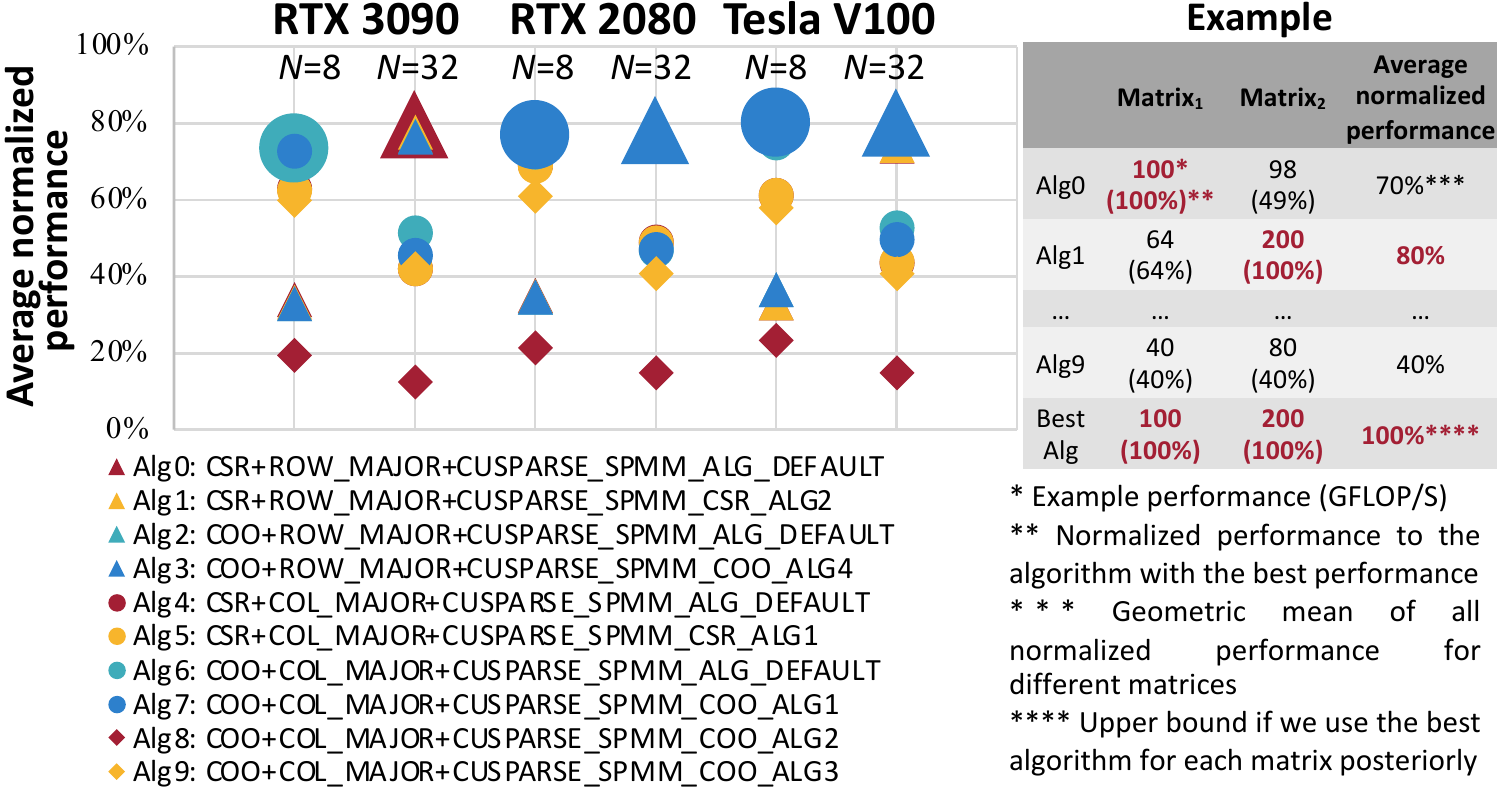}
    \caption{The left part shows the average normalized performance of different SpMM algorithms in cuSPARSE~\cite{cusparse} (to the algorithm with the best performance, using the geometric mean on the SuiteSparse~\cite{davis2011university}), $N$ is the width of the dense matrix. The right part shows an example of how the average normalized performance is calculated for 2 matrices.}
    %\vspace{-12pt}
    \label{fig:cusparsekernel}
\end{figure}

Sparsity is a key enabler of future Artificial Intelligence~\cite{low2012distributed, xiao2017tux2, han2015deep}. Sparsity enables fast and energy-efficient training and inference in various domains~\cite{eie, mobilenets, sanh2020movement, fedus2021switch}. However, sparse computation loses to its dense counterpart in terms of throughput, especially on throughput-oriented architectures like GPUs. In many emerging applications such as Graph Neural Networks (GNNs)\cite{gespmm}, recommendation~\cite{Asgari2021FAFNIRAS} and large language models~\cite{wang2019structured}, Sparse Matrix-Matrix Multiplication (SpMM) has served as fundamental components and dominated total execution time~\cite{gespmm, eie}. SpMM can be formulated by the multiplication between a sparse matrix and a dense matrix, exposing high potential parallelism and requires high throughput, thus many previous studies focus on accelerating SpMM using GPUs~\cite{aspt, jiang2020novel, gespmm, sputnik, wang2020sparsert}.

However, improving the performance of SpMM suffers from the challenge of handling the input dynamics, \textit{i.e.,} the performance is dependent on not only the static code optimizations, but also the dynamic input characteristics. We test the performance of 10 SpMM algorithms in NVIDIA cuSPARSE~\cite{cusparse} on different sparse matrices (956 matrices in the SuiteSparse~\cite{davis2011university} dataset) and different width of the dense matrix ($N=8$ and $N=32$). Figure~\ref{fig:cusparsekernel} shows performances of ten algorithms normalized to the best algorithm for each matrix. A single algorithm cannot always achieve the best performance on different input data, and the maximum performance loss with a single algorithm is >85\% (\textit{e.g.,} Alg8, $N=32$ on RTX 3090). Despite its importance and obvious difficulty, tackling input dynamics is missing from the focus of previous work. For example, GE-SpMM~\cite{gespmm} adopts techniques like thread coarsening and shared memory to optimize the performance when $N$ is large. However, GE-SpMM is slower than cuSPARSE when $N$ is small (e.g., running GraphSAGE~\cite{graphsage}, $N$=16, in GE-SpMM's paper). Another example is that the thread coarsening technique cannot always achieve benefit on different data (Figure 8 in GE-SpMM's paper).% (guyue): I think we need more example, simple statements

% \begin{figure}[!tp]
%     \centering
%     \includegraphics[width=0.25\textwidth]{figure/motivation-new.pdf}
%     \vspace{-15pt}
%     \caption{The scope of this work on SpMM problems compared with previous works~(e.g., TACO~\cite{taco}, Nisa \textit{et al.}~\cite{Nisa2018}, GE-SpMM~\cite{gespmm}, GraphBLAST~\cite{graphblast}). We introduce a novel design space for SpMM and an ML-based heuristic method to maximize the kernel performance of SpMM on GPUs.}
%     \vspace{-25pt}
%     \label{fig:motivation}
% \end{figure}

Observing that previous work limits their optimizations to some specific scenarios, we ask this question: how can we design a \textit{generally good} SpMM kernel for all kinds of scenarios? Our answer is that not a single design, but the \textit{careful composition and adaptive usage} of many designs, can well address input dynamics. Previous work has proposed many ad-hoc optimizations with limited rule-based selection models. In this paper, we seek to propose design-composition-guided new optimizations, and use data-driven ML models to make input-dynamic choices of implementations. We also solve the following challenges:
%\dl{Thus, to overcome the shortcoming that a single algorithm or specific optimizations cannot be adaptive to input dynamics, we consider the SpMM acceleration problem from a novel auto tuning perspective~\cite{chen2018tvm, graphit}. However, introducing auto tuning for SpMM needs to solve following challenges:}
\textbf{(1) Orthogonal design principles considering sparsity.} Optimizations in dense problems consider little of sparsity in SpMM. We need to extract orthogonal design principles considering sparsity to form algorithms for auto tuning. \textbf{(2) Nontrivial implementations in the algorithm space.} Combining techniques (\textit{e.g.,} workload balance for unbalanced matrices~\cite{graphblas} and parallel reduction to utilize idle GPU threads~\cite{bell2009implementing}) can bring benefits to specific inputs, while implementing them is nontrivial and missed in previous studies. \textbf{(3) Heuristic adaptability to input dynamics.} An efficient heuristic model is required to be adaptive to input dynamics. We make following contributions in this paper:

\squishlist

    \item \modified{\textbf{We propose three sparsity-unique design principles for SpMM on GPUs.} We decompose the mathematical formulation of the SpMM, with orthogonal design principles specially for sparse problems. We propose 8 algorithms according to 3 dimensions in the space, which not only cover previous designs, but also come up with new designs which are not studied.}
    
    \item \textbf{We propose nontrivial techniques to complete the algorithm space.} We propose several techniques complete missing algorithms in the space. For example, we introduce the conditional reduction to implement workload balance for imbalance matrices and parallel reduction to utilize idle GPU threads.
    
    \item \textbf{We design a data-aware kernel, \kernelname, to be adaptive to input dynamics.} \kernelname considers input dynamics, and dynamically selects a design with better performance for SpMM. \kernelname improves the performance of a static design from $<$70\% to $>$98\% (normalized to the best design on each dataset).
    
\squishend

DA-SpMM achieves \textbf{1.26$\times$}$\sim$\textbf{1.37$\times$} speedup on average compared with the best NVIDIA cuSPARSE algorithm, and brings up to \textbf{5.59$\times$} end-to-end speedup to applications like Graph Neural Networks.

%The following of this paper is organized as: Section~\ref{sec:relatedwork} introduces related works. Our three-dimension algorithm space with optimizations is shown in Section~\ref{sec:designspace}. Techniques for miss algorithm implementation are introduced in Section~\ref{sec:impl}. Our heuristic SpMM kernel, \kernelname, is detailed in Section~\ref{sec:kernel}. Extensive experimental results are shown in Section~\ref{sec:exp}. Section~\ref{sec:conclusion} concludes this paper.

\section{Related Works}\label{sec:relatedwork}

\subsection{Design Principles for Sparse}
TVM~\cite{chen2018tvm} uses affine-loop transformation which describes orthogonal design principles for dense tensor operations. Such a framework cannot handle sparse tensor operations with input-dependent loop boundaries. %In terms of sparse tensor operations, 
TACO~\cite{taco} proposes the sparse iterations space and a set of transformation primitives to support tuning. Unfortunately, TACO fails to consider some key techniques (\textit{e.g.,} parallel reduction and GPU shared memory) in their current framework, losing big room for optimization.

\subsection{SpMM Optimization Techniques}
% Under the existing GPU hardware architecture and data structure, how to accelerate the calculation of SpMMul is a very important issue in current research. The parallel programming model of GPU has its richness and complexity, such as thread arrangement and allocation, shared memory and cache model, warp scheduling and synchronization, etc. Therefore, researchers will start from these dimensions to think about more optimized hardware acceleration algorithms.

Bell \textit{et al.}~\cite{bell2009implementing} proposed a set of GPU SpMV implementations including Scalar-CSR and Vector-CSR, and the latter one can utilize idle GPU thread with a parallel reduction technique. GraphBLAS~\cite{graphblas} and GE-SpMM~\cite{gespmm} extend Scalar-CSR to SpMM scenario. CSR-Stream~\cite{greathouse2014efficient} and MergePath~\cite{merrill2016merge} handles workload balancing by introducing corresponding techniques, which work well for sparse matrices with the skewed row-length distribution. However, none of these works enable parallel reduction for SpMM and even combine parallel reduction with workload balance for skewed matrices on GPUs with high parallelism.

\subsection{Input-Adaptive Kernel Selection}%Data Characteristic Analysis for SpMMul}

Choi \textit{et al.}~\cite{choi2010model} searches for optimal format and different binning strategies. Such heuristics over-simplify input dynamics to one or a few features. Nisa \textit{et al.}~\cite{Nisa2018} explores various ML models for format selection of sparse tensor kernels. SpTFS~\cite{10.5555/3433701.3433724} explores deep learning to select a sparse tensor format. No prior work has studied machine learning techniques for general SpMM tuning.

\begin{figure}[!tp]
    \centering
    \includegraphics[width=0.45\textwidth]{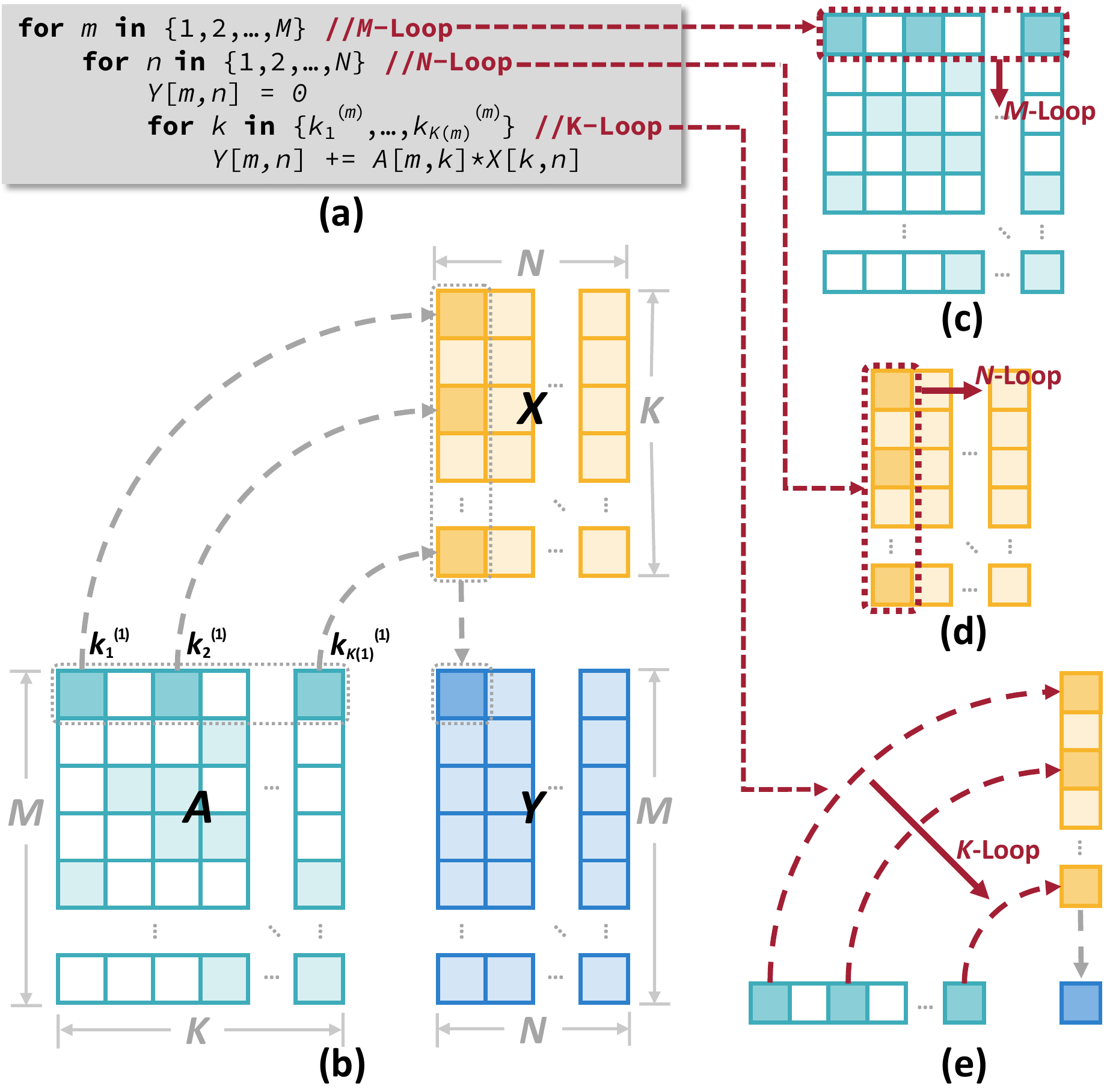}
    \caption{An overview of the SpMM. (a) Pseudo-code of SpMM with 3 loops. (b) Define SpMM as $A_{M\times K} \cdot X_{K\times N} = Y_{M\times N}$.  (c) $M$-loop: Traversing rows of $A$. (d) $N$-loop: Traversing columns of $X$. (e) $K$-loop: Traversing non-zeros in $A$ and corresponding elements in $X$.}
    \label{fig:overall}
\end{figure}

\section{Design Principles for SpMM on GPUs} \label{sec:designspace}

%This section introduces the algorithm space for SpMMul on GPUs from a novel three-loop perspective. The space is composed of three dimensions, which stem from the input dynamics of SpMMul design. We also propose corresponding optimization techniques for each dimension, and conclude previous designs in this space.

%\begin{figure}[h]
%    \centering 
%    \subfloat[\label{fig-dataflow}]{
%        \includegraphics[width=0.8\linewidth]{figure/dataflow.pdf}
%    }\\
%    \subfloat[\label{fig-design1}]{
%        \includegraphics[width=0.8\linewidth]{figure/designspace_1_balance.pdf}
%    }\\
%    \subfloat[\label{fig-design2}]{
%        \includegraphics[width=0.8\linewidth]{figure/designspace_2_vectorize.pdf}
%    }\\
%    \subfloat[\label{fig-design3}]{
%        \includegraphics[width=0.8\linewidth]{figure/designspace_3_layout.pdf}
%    }
%    \caption{Dataflow Graph and Design Space of SpMV/SpMM}
%\end{figure}

% This section introduces three new dimensions of the design space, that have not been summarized in previous work, but covers the input-dependent features of SpMV/SpMM design and we believe are what distinguishes SpMV/SpMM from dense workload on throughput-oriented architectures like GPUs. For each design dimension proposed, we first explain the design choice and show quantification on the SuiteSparse dataset that prove the feature does span a design dimension, i.e. reveals input-dependent trade-offs.

\subsection{Three-Loop for SpMM} \label{sec:designspace:designspace}

We explore the SpMM problem by digging into three loops of the typical SpMM implementation shown in Figure~\ref{fig:overall}(a). In general, we do not introduce detailed concepts of GPUs when explaining three loops. We use the concept of \textbf{worker} to represent parallel processing units (e.g., threads, warps, and etc.) on GPUs.

\begin{figure*}[!tp]
    \centering
    \begin{minipage}[t]{0.31\textwidth}
        \centering
        \includegraphics[width=\textwidth]{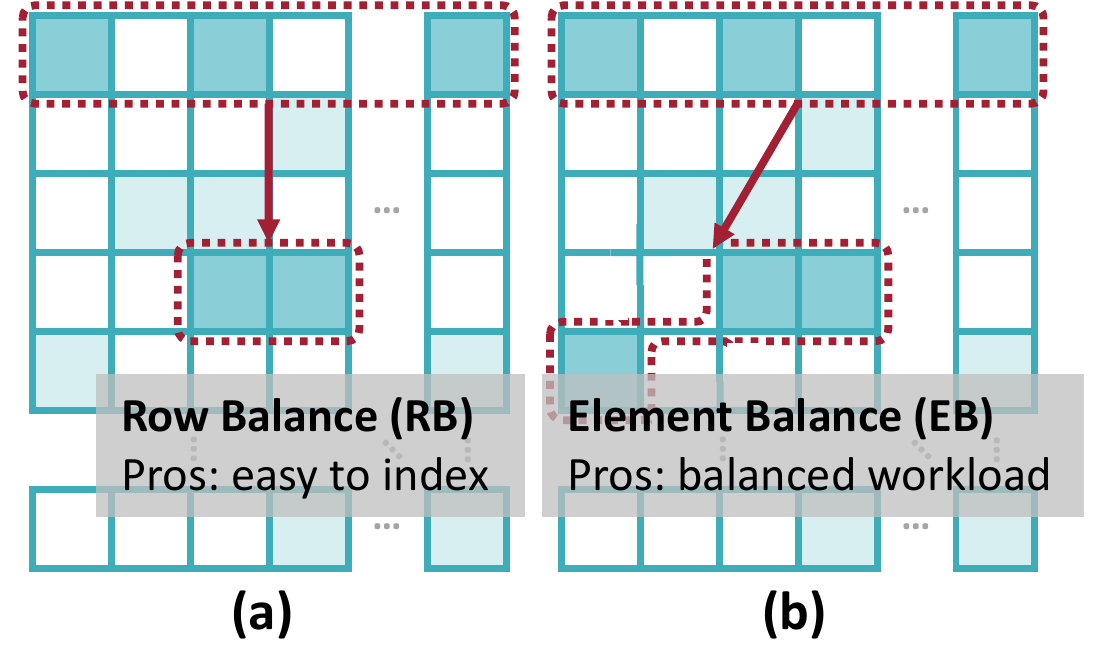}
        \caption{$M$-loop. (a) Row Balance (RB): assigning a row to a worker. (b) Element Balance (EB): assigning a given number of elements to a worker.}
        \label{fig:rb-eb}
    \end{minipage}
    \begin{minipage}[t]{0.02\textwidth}
        \centering
        \includegraphics[width=\textwidth]{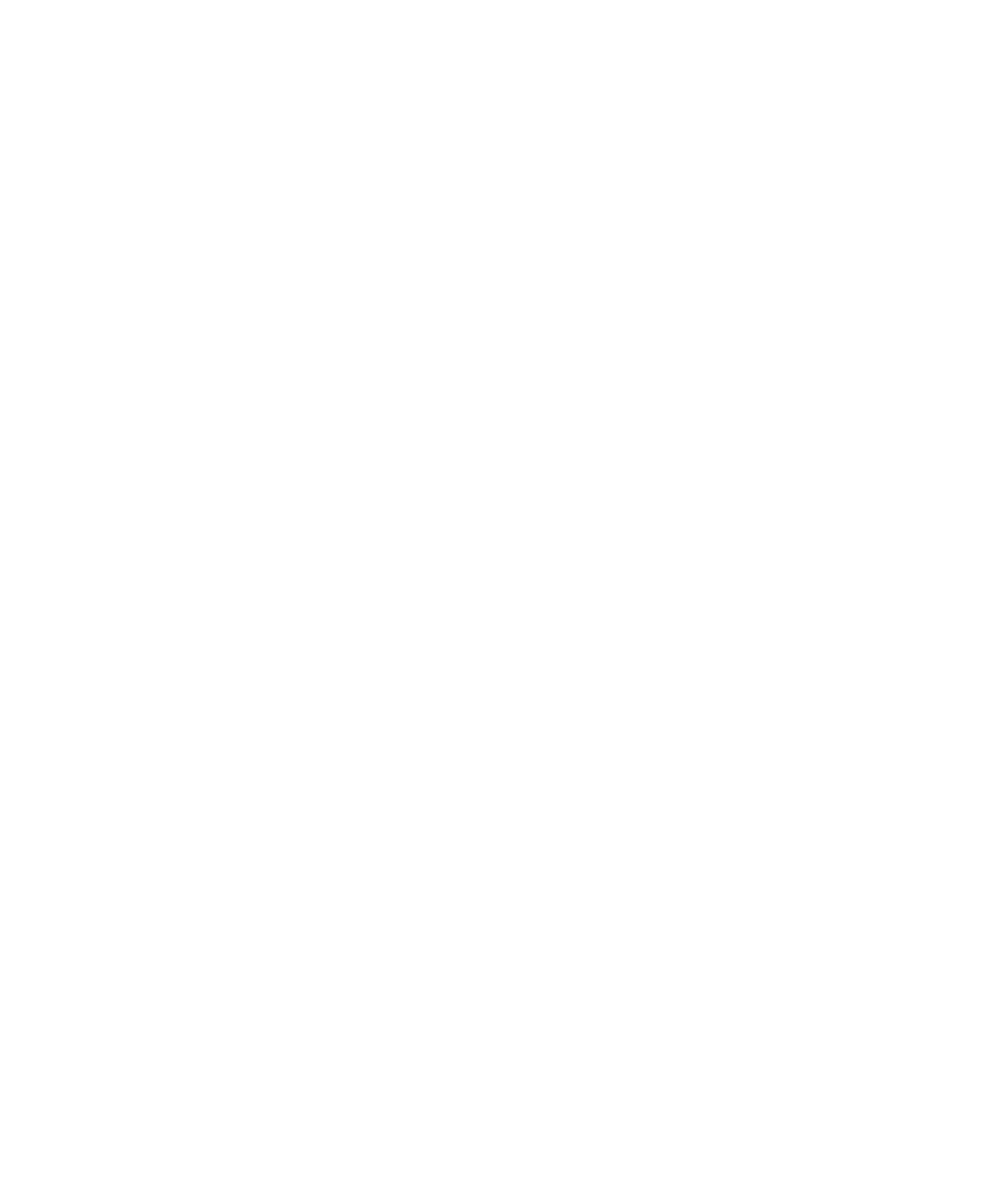}
    \end{minipage}
    \begin{minipage}[t]{0.31\textwidth}
        \centering
        \includegraphics[width=\textwidth]{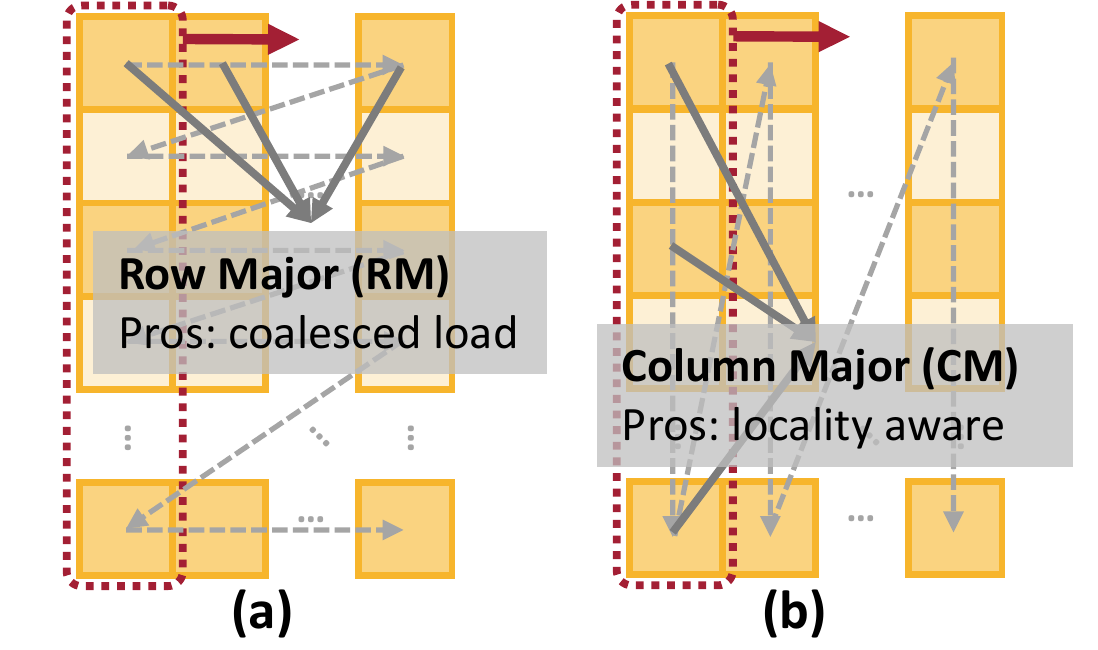}
        \caption{$N$-loop. (a) Row Major (RM): storing $X$ in the row-oriented order. (b) Column Major (CM): storing $X$ in the column-oriented order.}
        \label{fig:rm-cm}
    \end{minipage}
    \begin{minipage}[t]{0.02\textwidth}
        \centering
        \includegraphics[width=\textwidth]{figure/blank.pdf}
    \end{minipage}
    \begin{minipage}[t]{0.31\textwidth}
        \centering
        \includegraphics[width=\textwidth]{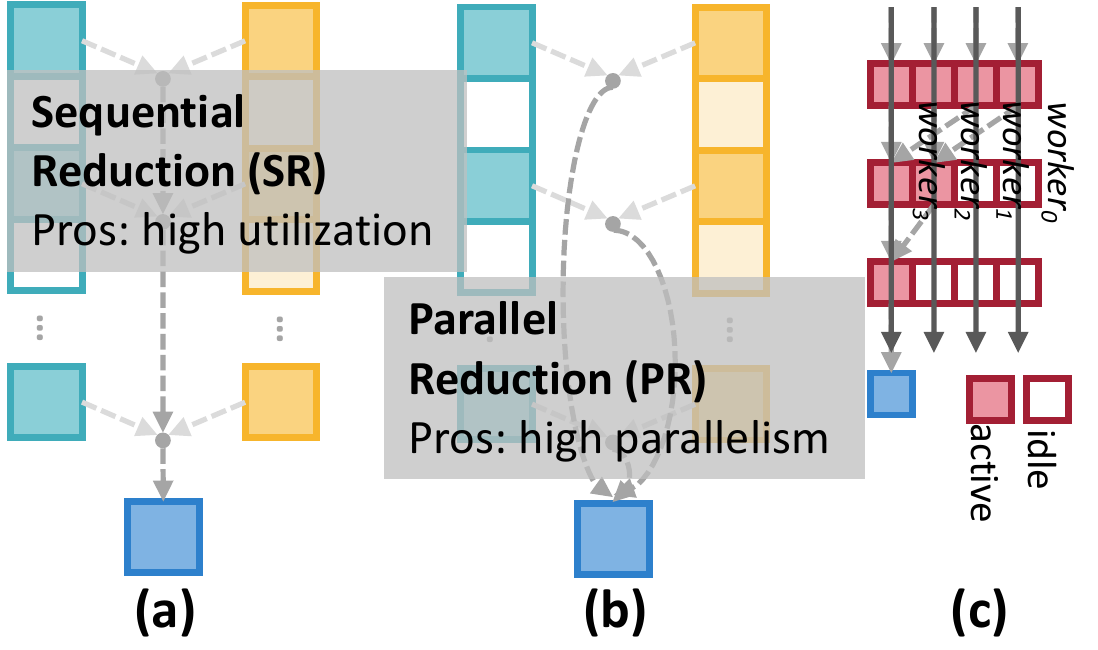}
        \caption{$K$-loop. (a) Sequential Reduction (SR): adding products sequentially.}
        \label{fig:sr-pr}
    \end{minipage}
    \end{figure*}

% \begin{figure}[!t]
%     \centering
%     \includegraphics[width=0.45\textwidth]{figure/rb-eb.pdf}
%     \vspace{-5pt}
%     \caption{Different design choices for the $M$-loop. (a) Row Balance (RB): assigning a row to a worker. (b) Element Balance (EB): assigning a given number of elements to a worker.}
%     \vspace{-5pt}
%     \label{fig:rb-eb}
% \end{figure}

% \begin{figure}[!t]
%     \centering
%     \includegraphics[width=0.45\textwidth]{figure/rb-eb-exp.pdf}
%     \vspace{-5pt}
%     \caption{Performance comparison between RB and EB. The EB method achieves better performance when non-zeros in different rows of the sparse matrix become more imbalance.}
%     \vspace{-5pt}
%     \label{fig:rb-eb-exp}
% \end{figure}

\subsection{$M$-Loop: Workload Balance} \label{sec:designspace:mloop}

\textbf{Intuition.} Consider the first loop, $M$-Loop, of SpMM shown in Figure~\ref{fig:overall}(a). $M$-Loop iterates over the rows of the sparse matrix $A$. For each row, the task is to multiply the non-zero elements with the dense elements they point to and accumulate results. If the $m$-th row contains $K_{(m)}$ non-zeros, $K_{(m)} \cdot N$ multiply-add operations are involved. The workload is thus specified for a given $m$ in $1 \sim M$.

\textbf{Design M1: Row Balance (RB).} A simple way for $M$-Loop unrolling and parallel worker assignment is to assign a given number of rows to a worker, which is commonly used in dense problems. Figure~\ref{fig:rb-eb}(a) shows a simple example of the RB method by assigning one row to a worker. By adopting this RB method, we can easily get the row index of each non-zero referring to the worker index. However, workloads of different workers are imbalanced because non-zeros vary in different rows.

\textbf{Design M2: Element Balance (EB).} Another way to unroll $M$-Loop is the fine-grained parallelism by assigning a given number of non-zeros in the sparse matrix $A$ to a worker, which is specially for sparse problems. Figure~\ref{fig:rb-eb}(b) shows an example of the EB method by evenly distributing non-zeros to different workers, where workloads are balanced. However, such a method introduces extra overheads including calculating the row index for a non-zero and partial sum aggregation among different workers in a row.

\textbf{Analysis.} Compared with the RB method, the EB method balances workloads by introducing extra calculations during runtime. Thus, the RB method is supposed to achieve a better performance when the number of non-zeros in different rows is balanced. On the contrary, the EB method tends to be used when workloads become more imbalanced. We later show the controlled experiment on how workload balance affects the choice of RB and EB in Section~\ref{sec:control}.%Obviously, the workload is related to the number of non-zeros in different rows, thus we can use the standard deviation of non-zeros of rows to represent the extent of imbalance.

% \begin{figure}[!t]
%     \centering
%     \includegraphics[width=0.45\textwidth]{figure/rm-cm.pdf}
%     \vspace{-5pt}
%     \caption{Different design choices for the $N$-loop. (a) Row Major (RM): storing $N$ row by row. (b) Column Major (CM): storing $N$ column by column.}
%     \vspace{-5pt}
%     \label{fig:rm-cm}
% \end{figure}

% \begin{figure}[!t]
%     \centering
%     \includegraphics[width=0.45\textwidth]{figure/rm-cm-exp.pdf}
%     \vspace{-5pt}
%     \caption{Performance comparison between RM and CM. RM achieves better performance when $N$ (width of the dense matrix) becomes larger because of benefits of coalesced loading.}
%     \vspace{-5pt}
%     \label{fig:rm-cm-exp}
% \end{figure}

\subsection{$N$-Loop: Dense Matrix Access Pattern}\label{sec:designspace:nloop}

\textbf{Intuition.} $N$-Loop iterates over the columns of the dense matrix $X$. Different workers process different columns of $X$ according to sparse elements in a given row of $A$ when $m$ is specified in Section~\ref{sec:designspace:mloop}. Note that workers access specific rows in parallel, thus the bandwidth efficiency caused by the memory access pattern is related to the data layout of the dense matrix $X$.

\textbf{Design N1: Row Major (RM).} We first consider the situation to store the dense matrix in a Row Major (RM) way, elements in a row of $X$ are stored in the contiguous memory space. Because different workers process the same rows (shown in Figure~\ref{fig:overall}), the RM layout leads to coalesced memory access pattern~\cite{gespmm}. However, such a data layout suffers from the poor locality for one worker because several rows are skipped, shown in Figure~\ref{fig:rm-cm}(a).

\textbf{Design N2: Column Major (CM).} Another data layout is to store $X$ using a Column Major (CM) order, shown in Figure~\ref{fig:rm-cm}(b). Compared with the RM layout, less memory space is skipped for a worker when processing two neighbor non-zeros in a column, which is beneficial to locality. However, parallel workers cannot benefit from coalesced memory access patterns because data in a row are not stored contiguously, shown in Figure~\ref{fig:rm-cm}(b).

\textbf{Analysis.} Note that the initial and final layout need to be aligned to the previous and next kernel in the application, thus often fixed, we have control on the intermediate matrix layout. When $N$ is small, the RM layout suffers from poor locality because several rows are skipped, while the CM major benefits from locality if the non-zero coordinates $k$'s appear close to each other. However, when $N$ is large, the RM layout shows advantage over CM because it exploits coalesced loading of a row of $X$, which increases the amount of effective data per request and improves bandwidth efficiency. We can easily infer that the parameter $N$ affects the performance of RM and CM for specific sparse matrices. We also show the controlled experiment on the choice of RM and CM in Section~\ref{sec:control}. %and tends to be better when the expected length of skipped elements in CM exceeds a physical threshold (e.g., cacheline size). We can easily infer that the expected length of skipped elements is related to the sparsity and row length ($N$) of the dense matrix $X$.

% \begin{figure}[!t]
%     \centering
%     \includegraphics[width=0.45\textwidth]{figure/sr-pr.pdf}
%     \vspace{-5pt}
%     \caption{Different design choices for the $K$-loop. (a) Sequential Reduction (SR): sequentially adding each product. (b) Parallel Reduction (PR): adding each product in parallel with primitives (e.g., warp reduction). (c) Schematic diagram of PR.}
%     \vspace{-5pt}
%     \label{fig:sr-pr}
% \end{figure}

% \begin{figure}[!t]
%     \centering
%     \includegraphics[width=0.45\textwidth]{figure/sr-pr-exp.pdf}
%     \vspace{-5pt}
%     \caption{Performance comparison between SR and PR. SR achieves better performance when the sparse matrix becomes larger, because it achieves a better utilization of each worker.}
%     \vspace{-5pt}
%     \label{fig:sr-pr-exp}
% \end{figure}

\subsection{$K$-Loop: Parallelism Efficiency} \label{sec:designspace:kloop}

\textbf{Intuition.} At last is the third loop, $K$-Loop. In this loop, non-zero elements of a row in $A$ and corresponding elements of a column in $X$ are traversed to calculate the inner product. Note that this loop requires a reduction operation among element pairs to get the sum, the efficiency of parallel workers is related to the way to implement such a reduction operation.

\textbf{Design K1: Sequential Reduction (SR).} A simple way to implement this reduction operation is to assign one worker to calculate the sum and perform the Sequential Reduction (SR) method, shown in Figure~\ref{fig:sr-pr}(a). The SR method achieves high utilization of workers because each worker is always busy when during reduction, which is often used for computation-bounded dense problems. However, such a method loses the potential parallelism existing in the reduction for sparse problems.% (e.g., parallel reduction shown in Figure~\ref{fig:sr-pr}(d)).

\textbf{Design K2: Parallel Reduction (PR).} Another implementation takes the potential parallelism existing in this reduction operation, as is shown in Figure~\ref{fig:sr-pr}(b). We call this method Parallel Reduction (PR), which has been used for sparse problems like Sparse Matrix-Vector Multiplication problems in~\cite{bell2009implementing}. Compared with SR, PR exploits the potential parallelism in reduction, and a typical implementation of PR is a merge-tree shown in Figure~\ref{fig:sr-pr}(d). However, PR suffers from poor utilization of workers because several workers are idle during the reduction, shown in Figure~\ref{fig:sr-pr}(c).

\textbf{Analysis.} The SR and PR methods show two ways of assigning the $K$-Loop to workers. In Figure~\ref{fig:sr-pr}(c), the PR method achieves higher parallelism while the efficiency/utilization of each parallel worker is lowered. When processing larger matrices, the parallelism lost by SR can be compensated by processing more reduction operations in parallel, and the performance tends to be better than the PR method due to higher utilization. On the contrary, PR tends to be better when processing smaller matrices (or GPUs with higher parallelism). The reason is that SR cannot fully utilize the parallelism provided by the GPU. Thus, the performance comparison between the SR and PR methods is closely related to the data size.

\begin{table}[!bp]
    \centering
    \small
    \caption{Summary of previous works in the algorithm space}
      \begin{tabular}{c|c|c|c|c|c|c|c|c}
      \hline
        $M$-Loop & RB    & RB    & RB    & RB    & EB    & EB    & EB    & EB \\
        $N$-Loop & RM    & RM    & CM    & CM    & RM    & RM    & CM    & CM \\
        $K$-Loop & SR    & PR    & SR    & PR    & SR    & PR    & SR    & PR \\
      \hline
        RowSplit\cite{graphblas} &  \checkmark     &       &       &       &       &       &       &  \\
        MergeSpMM\cite{graphblas} & \checkmark      &       &       &       &       &       &       &  \\
        ASpT\cite{aspt} &       &       &       &       &   \checkmark    &       &       &  \\
        GE-SpMM\cite{gespmm} &  \checkmark     &       &       &       &       &       &       &  \\
      \hline
      Ours &   \checkmark    &  \checkmark     &   \checkmark    &    \checkmark   &   \checkmark    &    \checkmark   &    \checkmark   & \checkmark \\
      \hline
      \end{tabular}%
      %\vspace{-15pt}
    \label{tab:designspacesummary}%
  \end{table}%

%\subsection{Implementations} \label{sec-designspace-impl}

% We also detailed all implementations with optimization methods (e.g., vectorized loading, and etc.) introduced in Section~\ref{sec:baseline} if appliable. 

% \begin{algorithm}[!tp]
%     \caption{Pseudo-code of SpMMul implementations}\label{alg:SpMMul}
%     \begin{algorithmic}[1]
%       \REQUIRE $A$ and $X$
%       \ENSURE $Y = A\cdot X$
%       \FOR {each $warp$ in parallel}
%       \STATE assign rows to $warp$
%       \FOR {each $warp$ with one row ($A[m,:]$) in parallel}
%       \FOR {each $X[:,n]$ in parallel}
%       \STATE load each $X[k,n]$
%       \STATE reduce\_sum ($A[m,k] \times X[k,n]$)
%       \ENDFOR
%       \ENDFOR
%       \ENDFOR
%       \FOR {each $Y[m,:]$ in parallel}
%       \STATE $Y[m,:]$ = atomic\_add (related $warps$)
%       \ENDFOR
%       \RETURN $Y$
%     \end{algorithmic}
%   \end{algorithm}

%   \begin{figure}[!bp]
%       \centering
%       \includegraphics[width=0.48\textwidth]{figure/ratio.pdf}
%       \vspace{-15pt}
%       \caption{Ratio of each design that achieves the best performance on SpMMul ($N$ is from 1 to 128).}
%       \label{fig:ratio}
%   \end{figure}

\subsection{Comparison with Previous Works}

Based on the novel three-loop model, we summarize previous SpMM designs in Table~\ref{tab:designspacesummary}. By decomposing the mathematical formulation of SpMM, combining design principles is capable of covering all previous designs. Moreover, we can also find new designs for the SpMM (\textit{e.g.,} RB+RM+PR.). We implement all 8 designs based on these design principles, which are further used for performance tuning in later sections.

\begin{figure}[!t]
    \centering
    \includegraphics[width=0.48\textwidth]{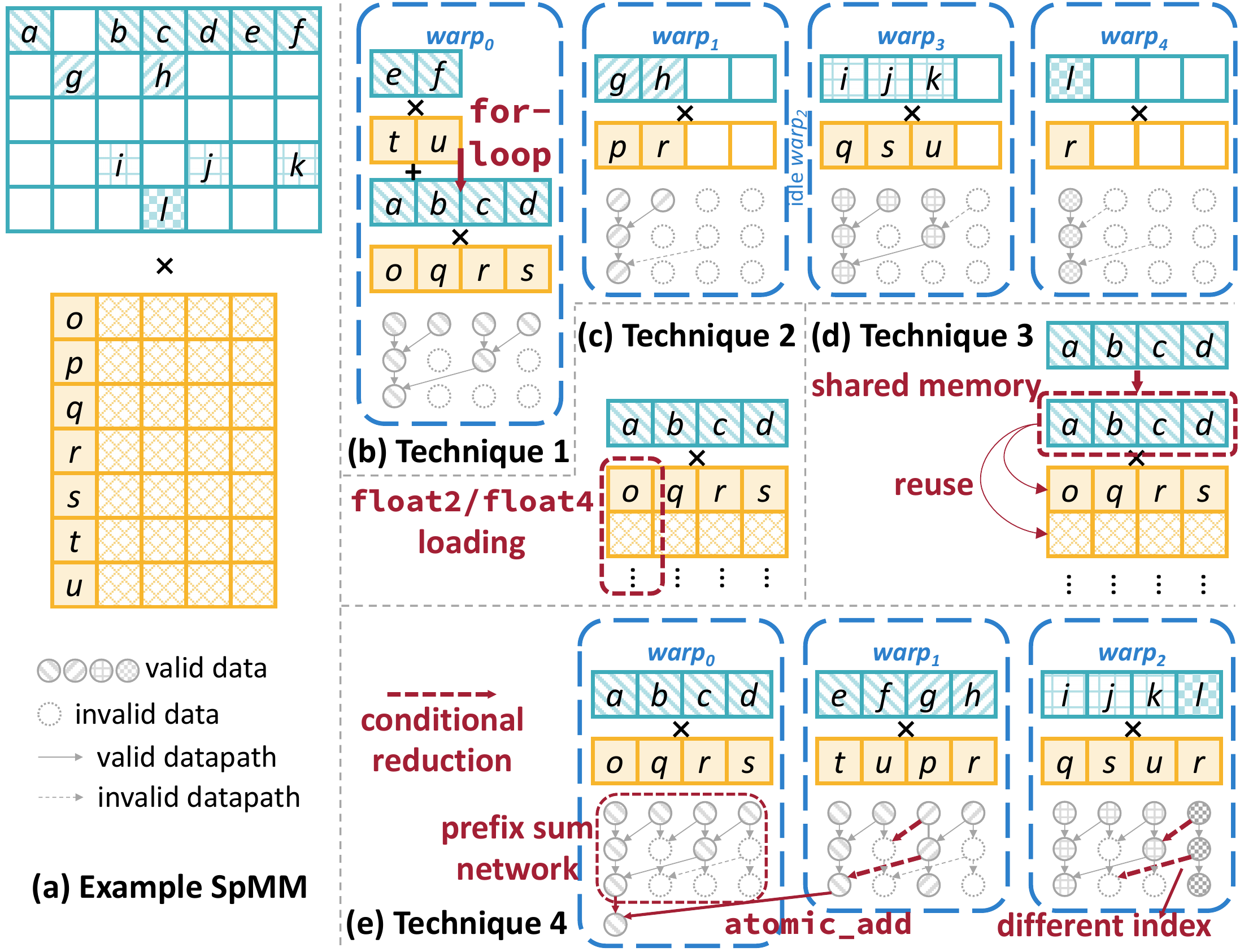}
    \caption{Implementations of different SpMM algorithms under the algorithm space with proposed techniques, number of threads in a warp is 4. (a) An example of SpMM with legends in latter subfigures. (b) Using \texttt{for-loop} to reduce $>4$ elements for RB. (c) Using \texttt{float2/float4} to load multiple dense elements for RM. (d) Using shared memory to cache sparse elements for reuse. (e) Using a prefix sum network with conditional reduction for EB.}
    \label{fig:impl}
\end{figure}

\section{Implementations} \label{sec:impl}

As mentioned in Table~\ref{tab:designspacesummary}, previous studies on SpMM only explore both RB and EB methods for the $M$-Loop, while CM for the $N$-Loop and PR for the $K$-Loop is lacking. Implementing CM is trivial derived from the implementation of the corresponding RM design by storing the dense matrix in a column-oriented order. Thus, we can easily get RB+CM+SR and EB+CM+SR from current implementations.

\textbf{RB+RM+PR}. The main challenges of implementing PR under RB include: (1) The number of elements to perform the reduction may be larger than the number of threads in a GPU warp, thus the reduction cannot be processed within one warp instruction. (2) Different columns require same non-zeros in a row of the sparse matrix, leading to redundant data loading. Figure~\ref{fig:impl}(a) shows an example of performing reduction on 5 sparse rows with 6/2/0/3/1 non-zeros each row, we set the number of threads in a warp to 4. We utilize a \textbf{\texttt{for-loop} (Technique 1)} in Figure~\ref{fig:impl}(b), for the first challenge to enable reduction of $>$4 (number of threads in a warp) elements (\textit{e.g.,} the first row). For the second challenge, we utilize \textbf{\texttt{float2/float4} (Technique 2)}, shown in Figure~\ref{fig:impl}(c), primitive to load multiple elements in consecutive columns in the dense matrix, and these columns are processed by a warp. However, the \texttt{float2/float4} primitive only enables us to process at most 4 columns within a warp. In order to reuse data for large $N$ (columns in the dense matrix), we further utilize \textbf{shared memory (Technique 3)}, shown in Figure~\ref{fig:impl}(d), to cache sparse elements for different columns. Thus, these elements can be reused for multiple columns.

\textbf{EB+RM+PR}. The main challenges of implementing PR under EB include: (1) A warp needs to perform reduction on different rows for the same number of elements in the sparse matrix. (2) The same challenge as the second challenge in RB+RM+PR, thus \textbf{Technique 2/3} can still be used. Inspired by the prefix sum network, we propose the \textbf{conditional reduction (Technique 4)}, shown in Figure~\ref{fig:impl}(e), method to solve the first challenge. Besides the partial reduction result, a thread records a \texttt{index} flag to distinguish the row. For each reduction step, the datapath of the prefix sum network is activated if two threads share the same \texttt{index} flag. The outputs of such a conditional prefix sum network are valid for the first (leftest) thread or the thread with different \texttt{index} compared with the thread on its left. For rows assigned to different warps, an \texttt{atomic\_add} operation is performed get the final reduction result.

\textbf{RB+CM+PR}. The only difference to RB+RM+PR is that the dense matrix is stored in a column-oriented order, thus \textbf{Technique 2} cannot be used because elements in a row are not stored consecutively, while \textbf{Technique 1/3} are applied.

\textbf{EB+CM+PR}. Similar to RB+CM+PR, \textbf{Technique 2} is not applied, while \textbf{Technique 3/4} in EB+RM+PR are used.

\begin{figure}[!tp]
    \centering
    \includegraphics[width=0.48\textwidth]{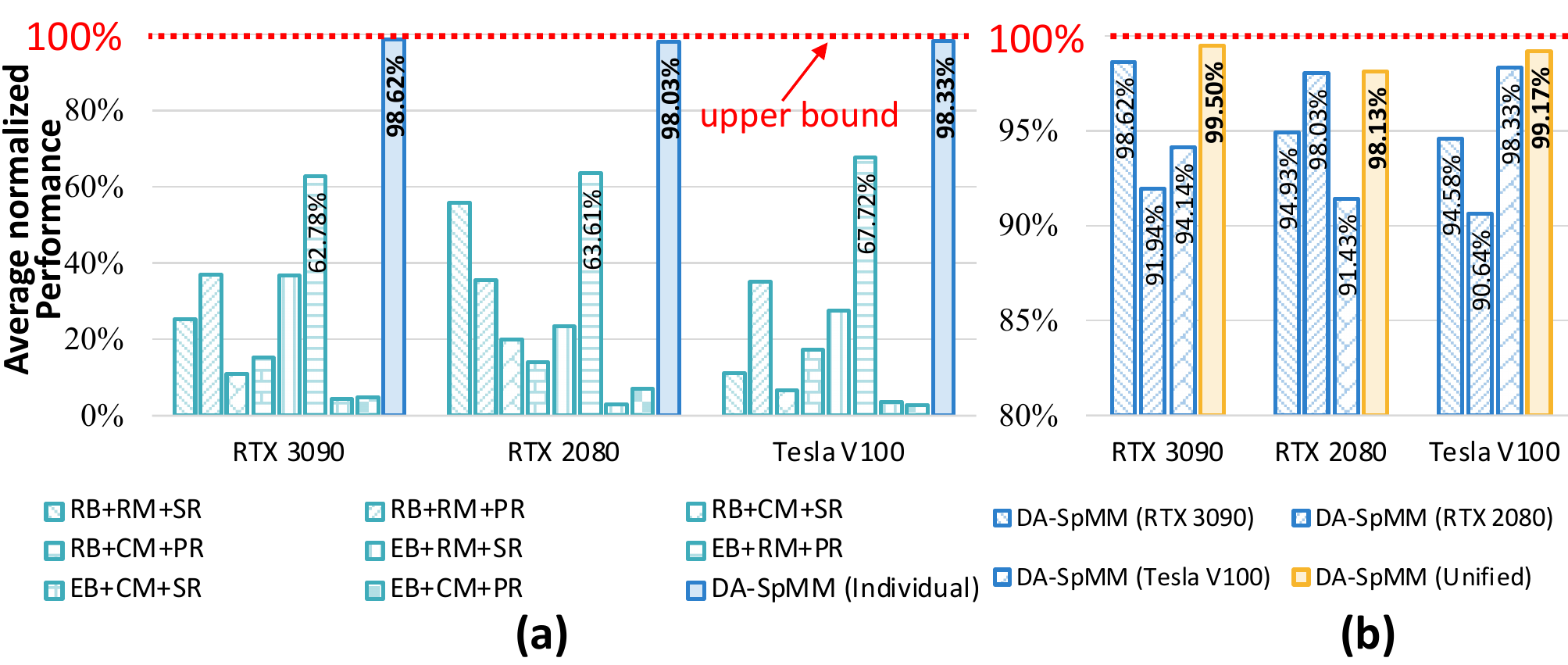}
    \caption{Average normalized performance of the heuristic DA-SpMM kernel. (a) \kernelname based on features of input data against 8 static designs proposed in Section~\ref{sec:designspace}. (b) A unified \kernelname kernel for different GPUs.}
    \label{fig:heuristickernel}
 \end{figure}

\section{Heuristic Kernels}\label{sec:kernel}

\subsection{Heuristic Adaptability}

We get 8 optimized SpMM algorithms in Section~\ref{sec:impl}, and we also analyze that characteristics of the input data affect the performance of implementations. Then, we further test the performance of all these 8 designs on 3 different GPUs with 956 matrices in the SuiteSparse dataset (with detail setups in Section~\ref{sec:exp:setup}). The average normalized performance (the meaning is the same as that in Figure~\ref{fig:cusparsekernel}) is shown in Figure~\ref{fig:heuristickernel}(a). We can get the same conclusion that a single algorithm cannot always achieve the best performance on different input data, and a single algorithm with the best average normalized performance is $<$70\%.

Thus, to be adaptive to input dynamics and achieve a better performance, auto tuning has been proved as an effective method. We leverage the gradient boosting framework LightGBM~\cite{lightgbm} to train a prediction model to select the best design from the 8 different designs. We select a collection of features extracted from both the sparse matrix and the dense matrix as the input vector to our model, as in Table~\ref{tab:feature}.%We also choose hardware features (e.g., size of L2 cache and number of CUDA cores) as input to show how hardware parameters affect the performance.

\begin{table}[!h]
    \centering
    \caption{Extracted features in heuristic kernels.}
    \small
      \begin{tabular}{c|l}
      \hline
      Feature & Meaning \\
      \hline
        \textit{nnz} & Number of non-zeros in the sparse matrix \\
        \textit{mat\_size} & Size of the sparse matrix \\
        \textit{std\_row} & Standard deviation of non-zeros in rows \\
        \textit{N} & Width of the dense matrix \\
      \hline
      \end{tabular}%
    \label{tab:feature}%
  \end{table}%

%The performance is shown in Figure~\ref{fig:ratio}. We can get a similar conclusion as the performance of different cuSPARSE kernels in Figure~\ref{fig:cusparsekernel}, one single implementation cannot achieve the best performance on different datasets.

%Thus, to be adaptive to input dynamics and achieve a better performance, auto tuning has been proved as an effective method. Among auto tuning solutions, Machine Learning (ML) techniques are commonly used for regression and classification problems. Compared with other ML methods, gradient boosting is characterized by synthesizing weak prediction models (usually decision trees), to derive a strong prediction model. 

% \begin{table}[!h]
% %\vspace{-10pt}
% \caption{Extracted features in heuristic kernels.}
% \small
% \centering
% \label{table:feature}
% \begin{tabular}{c|l}
% \hline
% Feature & Meaning \\ \hline
% \textit{nnz} & number of non-zeros in the sparse matrix \\ 
% \textit{mat\_size} & size of the sparse matrix \\ 
% \textit{avg\_row} & average number of non-zeros in rows \\ 
% \textit{avg\_col} & average number of non-zeros in columns \\ 
% \textit{dev\_row} & standard deviation of non-zeros in rows \\ 
% \textit{dev\_col} & standard deviation of non-zeros in columns \\ 
% \textit{N} & width of the dense matrix \\ \hline
% \end{tabular}
% %\vspace{-10pt}
% \end{table}

% Table generated by Excel2LaTeX from sheet 'Sheet1'

\begin{figure*}[!tp]
    \centering
    \includegraphics[width=\textwidth]{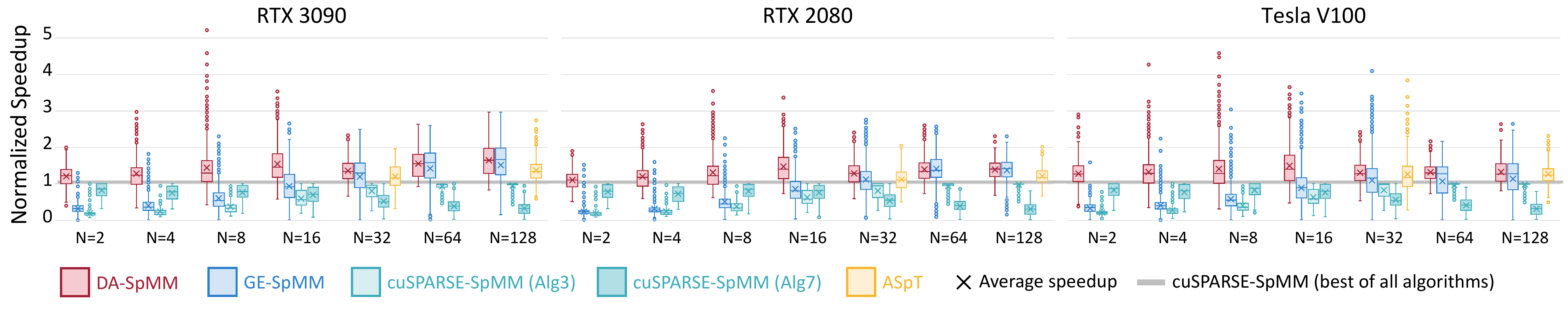}
    \caption{SpMM performance comparison among our DA-SpMM, GE-SpMM~\cite{gespmm}, and ASpT~\cite{aspt}, and cuSPARSE~\cite{cusparse}.}
    \label{fig:overall-result}
\end{figure*}
 
\subsection{Evaluation}

\subsubsection{Individual Model.}
We also use the average normalized performance as the criterion to evaluate the performance of our heuristic kernel, \kernelname. Here we train three individual models for each GPU. We set the ratio of train, validation, and test data to 40\%, 10\%, and 50\%, respectively. As is shown in Figure~\ref{fig:heuristickernel}(a), \kernelname achieves \textbf{98.62\%}, \textbf{98.03\%}, and \textbf{98.33\%} normalized performance on 3 GPUs (an average of \textbf{98.33\%}), respectively. In contrast, the best normalized performance of one static design is lower than 70\%.

%Thus, \kernelname solves the problem that the static kernel optimization cannot be adaptive to the input dynamics.

\subsubsection{Unified Model.} We further test the performance of migrating the model trained by one GPU to another, leading to an average of 5.38\% performance loss (\textit{e.g.,} using \kernelname model trained by RTX 2080, \textit{i.e.,} DA-SpMM (RTX 2080), for Tesla V100 degrades the performance from 98.33\% to 90.64\%). To be adaptive to different GPUs, we further distinct different GPUs to train one unified LightGBM model. The result shows that, a unified \kernelname can not only solve the performance loss of migrating the model, it also further achieves the better performance, \textbf{99.50\%}, \textbf{98.13\%}, and \textbf{99.17\%} on 3 GPUs (an average of \textbf{98.93\%}).

\subsection{Heuristic Kernel Conclusion}

By introducing the heuristic model to dynamically select kernels with better performance, \kernelname is adaptive to input dynamics. With only data features for model training, \kernelname improves the normalized performance from less than 70\% to over 98\% for different GPUs. By introducing hardware features, \kernelname further improves the performance with a unified heuristic model.

\section{Experimental Results}\label{sec:exp}

\subsection{Experimental Setup}\label{sec:exp:setup}

\subsubsection{Environments.}\label{sec:exp:setup:env}

We use three different GPUs set up our experimental environments for previous and this sections:

NVIDIA \textbf{RTX 3090.} Compute Capability 8.6 (68 Ampere SMs at 1.395 GHz, 24 GB GDDR6x, 936 GB/s bandwidth).%Host CPU: AMD Ryzen Threadripper 3970X (32 cores).

NVIDIA \textbf{RTX 2080.} Compute Capability 7.5 (46 Turing SMs at 1.515 GHz, 8 GB GDDR6, 448 GB/s bandwidth).%Host CPU: Intel Core i7-9700K (8 cores).

NVIDIA \textbf{Tesla V100.} Compute Capability 7.0 (80 Volta SMs at 1.370 GHz, 16 GB HBM2, 900 GB/s bandwidth).%Host CPU: Intel Xeon CPU E5-2698 v4 (20 cores).

\subsubsection{Baselines.}

\textbf{GE-SpMM}~\cite{gespmm} is a state-of-the-art open-source SpMM design on GPUs, it achieves better performance against other open-source design like GraphBLAST~\cite{graphblast}.

\textbf{ASpT}~\cite{aspt} adopts a matrix tiling preprocessing step to process dense and sparse sub-matrix separately during runtime. We only count the execution time of ASpT while excluding the preprocessing time. Moreover, ASpT only supports $N=32$ and $N=128$ cases.

\textbf{NVIDIA cuSPARSE}~\cite{cusparse} is a vendor library by NVIDIA. We use the routine \texttt{cusparseSpMM} for SpMM, we use version 11.2.

All results are normalized to the best cuSPARSE algorithm for each matrix. All codes are compiled using \texttt{nvcc} 11.2 with -\texttt{O3} flag.

\subsubsection{Datasets.}\label{sec:exp:setup:dataset}

We use SuiteSparse collection~\cite{davis2011university} with 956 matrices (We use the script of ASpT~\cite{aspt} to select these matrices which removes isomorphic one to enable diversity) to test performance. 

\subsection{Performance of Heuristic Kernel}

We compare the DA-SpMM design with GE-SpMM~\cite{gespmm}, ASpT~\cite{aspt}, and two cuSPARSE~\cite{cusparse} SpMM algorithms (Alg3 and Alg7 in Figure~\ref{fig:cusparsekernel}, which are two best static algorithms). All results are normalized to the best cuSPARSE SpMM algorithm on each matrix, shown in Figure~\ref{fig:overall-result}. We can get following conclusions:

%\squishlist

    %\item \textbf{Against static cuSPARSE.} For different $N$ from 2 to 128, \kernelname achieves an average speedup of \textbf{2.60$\times$} to \textbf{2.96$\times$}, \textbf{2.42$\times$} to \textbf{2.66$\times$}, \textbf{2.36$\times$} to \textbf{2.57$\times$} on three GPUs, respectively.

    \textbf{Against the static cuSPARSE algorithm.} For different $N$ from 2 to 128, \kernelname achieves an average speedup of up to 2.96$\times$, 2.66$\times$, 2.57$\times$ on three GPUs, respectively.

    \textbf{Against different cuSPARSE algorithms with the best performance on each matrix.} For different $N$ from 2 to 128, DA-SpMM achieves an average speedup of 1.37$\times$, 1.26$\times$, and 1.30$\times$ on three GPUs, respectively. For large $N$ (e.g., $N=128$), \kernelname achieves higher speedup of 1.60$\times$, 1.36$\times$, and 1.30$\times$, respectively. 
    
    \textbf{Against ASpT.} For $N=32$, DA-SpMM achieves an average speedup of 1.14$\times$, 1.14$\times$, and 1.06$\times$, respectively. For $N=128$, the speedup is 1.20$\times$, 1.14$\times$, and 1.05$\times$. The preprocessing time of ASpT is not counted.

    \textbf{Against GE-SpMM.} DA-SpMM achieves significant improvements for small $N$ (4.99$\times$, 5.23$\times$, and 5.40$\times$ for $N=2$). When $N$ becomes larger, the performance of GE-SpMM improves. For $N=128$, the speedup against GE-SpMM is 1.29$\times$, 1.04$\times$, and 1.59$\times$ on three GPUs, respectively. When $N$ becomes larger, DA-SpMM prefers to select RM for $N$-loop and SR for $K$-loop in most cases. These two selections are the same with the design in GE-SpMM, which means that GE-SpMM does reach the optimal design point for these cases, except for the balance design in the $M$-loop.

%\squishend

\subsection{Controlled Experiments}\label{sec:control}

As we have analysed in Section~\ref{sec:designspace}, the choice of different algorithm for each loop depends on input dynamics. Thus, we synthesize matrices and conduct controlled experiments.

\subsubsection{RB-EB} We take three sparse matrices with identical size and sparsity, but different $std\_nnz$, \textit{i.e.,} the standard deviation of the number of non-zeros in each row. The matrices are synthesized R-MAT graphs ~\cite{chakrabarti2004r} by tuning parameters related to element distribution. The result is shown in Figure~\ref{fig:control}(a). The EB method exceeds the RB method as the standard deviation of non-zeros in each row grows, because the imbalance among different workers becomes the main factor for the performance difference.

\subsubsection{RM-CM} We synthesize one same sparse matrix with three different dense matrices, with R-MAT method~\cite{chakrabarti2004r}. We vary $N$ for the dense matrix. The result is shown in Figure~\ref{fig:control}(b). The RM layout achieves higher performance against the CM layout when $N$ gets larger. The reason is that larger $N$ leads to coalesced load for different workers. In contrast, the CM implementation always needs to skip elements in the dense vector, thus the effective data per request is inherently limited.%and worse locality on the column dimension because more elements in a row are skipped.

\subsubsection{SR-PR} 
\begin{figure}[!tp]
    \centering
    %\vspace{-10pt}
    \includegraphics[width=0.48\textwidth]{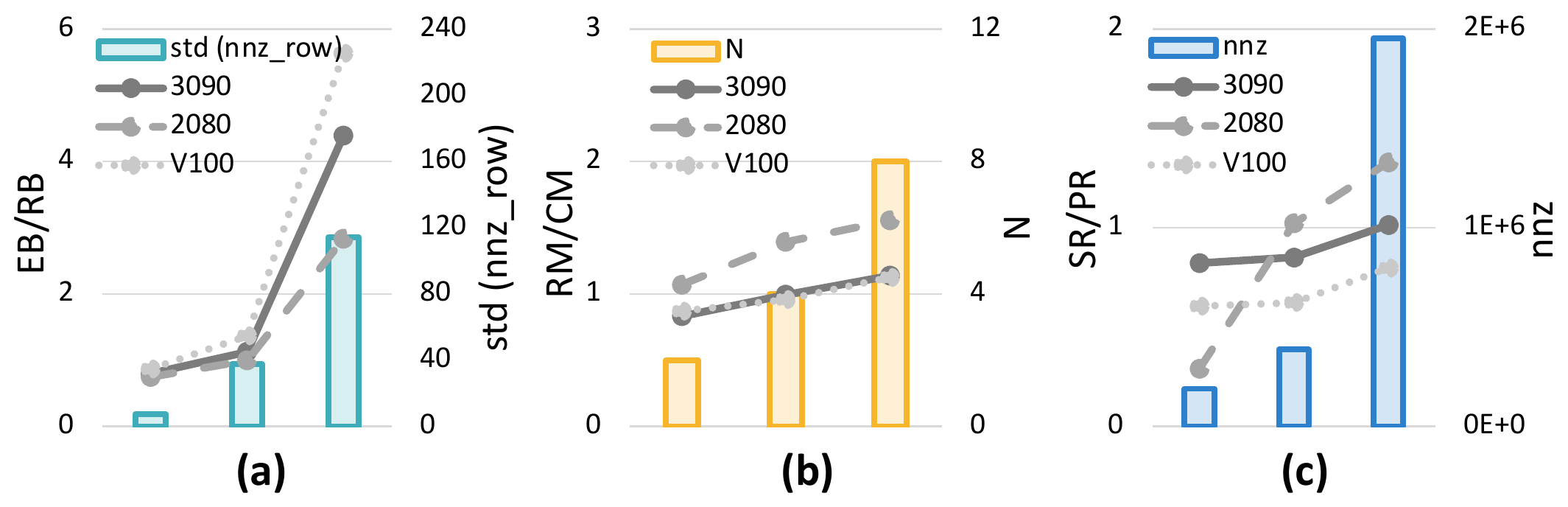}
    \caption{Controlled experiments to show how input data dynamics affect choice of different algorithms for each loop. (a) RB-EB: workload balance. (b) RM-CM: data access pattern. (c) SR-PR: total workloads.}
    \label{fig:control}
 \end{figure}

We synthesize three matrices with the same sparsity and non-zero distribution, but different non-zeros, with R-MAT method~\cite{chakrabarti2004r}. The bars in Figure~\ref{fig:control}(c) display non-zeros of each matrix, and the lines plot the relative performance of SR against PR. When the matrix is small, the PR method is better than the SR method because it saturates parallelism, but SR gains advantages when the matrix size gets larger because the utilization of one worker becomes the main factor for the performance.

\subsection{Performance of Heuristic Kernel in GCNs}

\begin{figure}[!tp]
    \centering
    \vspace{-5pt}
    \includegraphics[width=0.48\textwidth]{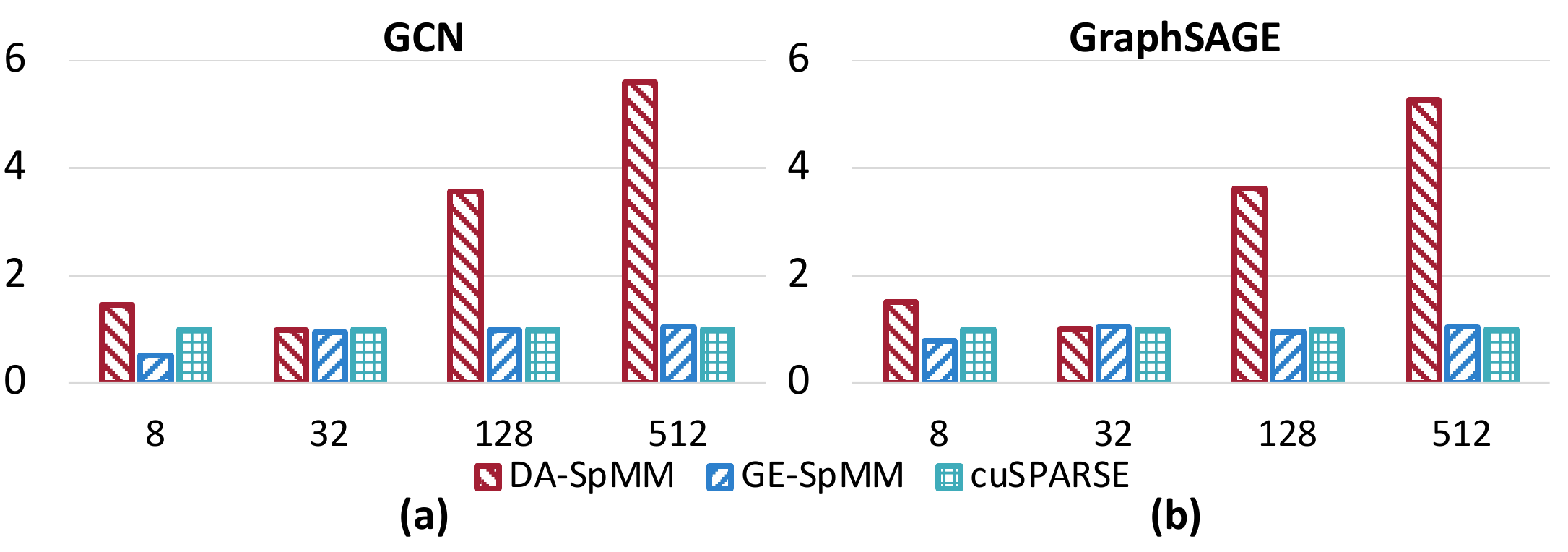}
    \caption{Performance comparison for GNNs. X-axis: feature length. Y-axis: Normalized speedup (to the original framework using cuSPARSE). (a) GCN. (b) GraphSAGE.}
    \label{fig:gnn}
 \end{figure}

We integrate DA-SpMMul into the Graph Neural Network framework, DGL~\cite{dgl}, to compare the end-to-end performance on RTX 3090 with GCN~\cite{kipf2016semi} and GraphSAGE~\cite{graphsage} models. We test the inference speedup with on the Reddit dataset~\cite{reddit}, in Figure~\ref{fig:gnn}. By using DA-SpMM, we achieve up to \textbf{5.59$\times$} and \textbf{5.27$\times$} speedup against the original DGL framework on two GNN models, respectively.

\section{Conclusions}\label{sec:conclusion}

In this paper, we present a thorough study of Sparse Matrix-Matrix multiplication (SpMM) on GPUs. We consider the SpMM acceleration problem from a novel auto-tuning perspective to be adaptive to input dynamics. We first propose a novel three-loop model with orthogonal design principles specially for sparse problems. We also propose several techniques to implement new designs which are not previously studied. We further propose \kernelname to adaptively optimize code and achieve a better performance considering input dynamics. Evaluations show that \kernelname is the fastest among commercial and academic designs: DA-SpMM achieves an average of \textbf{1.26$\times$}$\sim$\textbf{1.37$\times$} speedup compared with the best NVIDIA cuSPARSE algorithm, and brings up to \textbf{5.59$\times$} end-to-end speedup to applications like Graph Neural Networks. Our methodology of composing design principles into an algorithm space and using heuristic models for algorithm selection can be extended to further studies on sparse acceleration problems.

\section{Acknowledgements}
The design in this paper will be included in our dgSPARSE project\footnote{The dgSPARSE project (\url{https://dgsparse.github.io/}) is an open source project for fast and efficient graph processing on GPUs. Currently the project provides three solutions: (1) \textbf{dgSPARSE Library} is a high performance GPU library for sparse operator acceleration (\textit{e.g.,} SpMM, SDDMM, and etc.) (2) \textbf{dgSPARSE Wrapper} provides the compatible interfaces to upper layer frameworks, and algorithms can benefit from different accelerated sparse operators without modifying codes. (3) \textbf{dgNN Library} provides high performance GNN layers for various GNN models (\textit{e.g.,} GAT, EdgeConv, and etc).}. 

%%
%% The next two lines define the bibliography style to be used, and
%% the bibliography file.

\bibliographystyle{unsrt}
\bibliography{ref}

%%
%% If your work has an appendix, this is the place to put it.

\end{document}